# Local majority-with-inertia rule can explain global consensus dynamics in a network coordination game

Felix Gaisbauer[1,2,*], Ariana Strandburg-Peshkin[3,4,5], and Helge Giese[2,4]

[1]Graduate School of Decision Sciences, University of Konstanz, Germany

[2]Department of Psychology, University of Konstanz, Germany

[3]Department of Biology, University of Konstanz, Germany

[4]Centre for the Advanced Study of Collective Behaviour, University of Konstanz, Germany

[5]Department for the Ecology of Animal Societies, Max Planck Institute of Animal Behaviour, Germany

* Correspondence and requests for materials should be addressed to F.G.

## ABSTRACT

We study how groups reach consensus by varying communication network structure and individual incentives. In 342 networks of seven individuals, single opinionated "leaders" can drive decision outcomes, but do not accelerate consensus formation, whereas conflicting opinions slow consensus. While networks with more links reach consensus faster, this advantage disappears under conflict. Unopinionated individuals make choices consistent with a local majority rule combined with "inertia" favouring their previous choice, while opinionated individuals favour their preferred option but yield under high peer or time pressure. Simulations show these individual rules can account for group patterns, and allow rapid consensus while preventing deadlocks.

**Keywords:** collective decision-making, group consensus, communication network, majority rule, social influence

## Introduction

Coming to consensus on a group decision is commonly required in both human societies (Boehm, 1996; Davis et al., 1976; Kerr and Tindale, 2004) and animal groups (Dyer et al., 2009; Smith et al., 2016; Strandburg-Peshkin et al., 2015). Consensus decisions are decisions where members of a group—sometimes also called a jury, team, crowd, or swarm (Hertwig, 2012)—choose between two or more mutually exclusive outcomes while seeking to maintain group cohesion (Conradt and Roper, 2005). *Shared* consensus decisions, where multiple members rather than a single individual contribute to the decision outcome, can produce less extreme and therefore more beneficial outcomes (Conradt and Roper, 2005, 2003; Hastie and Kameda, 2005; List, 2004; Simons, 2004). However, individuals in groups are often constrained to rely on *local* rather than *global* communication (Conradt and Roper, 2005), with individuals interacting directly with only a subset of group members. Such constraints on the *communication network* of a group may produce a challenge to effective coordination (Lerman et al., 2016; Stewart et al., 2019). The majority of studies on consensus decision-making in humans have ignored the underlying communication structures by (implicitly) assuming *global* communication where all group members can communicate directly (Kearns et al., 2009; but see Stewart et al., 2019). More constrained communication networks, however, can vary dramatically in their propensity to spread information (Barkoczi and Galesic, 2016; Kearns, 2006; Kearns et al., 2009; Watts and Strogatz, 1998) and behaviour (Centola, 2010; Christakis and Fowler, 2007; Giese et al., 2017), potentially influencing both the speed and the outcomes of group decisions (Lerman et al., 2016; Stewart et al., 2019).

Generally, the speed of coming to consensus should be dependent on the density, i.e. the total number of communication links in a network: The more communication exists between individuals (up to global communication), the easier it should be to coordinate on a group's outcome (e.g., Centola, 2010). Beyond the number of communication links, the literature furthermore discusses specific link structures that may facilitate or impede group consensus (Burt et al., 2021; Centola, 2010; Freeman et al., 1979; Kearns, 2006; Kearns et al., 2009) with a focus on two types of network topologies: Studies on leadership and coordination propose that hierarchical structures that decrease the overall distance between individuals in a group and small-world designs are particularly effective to promote group coordination (Burt et al., 2021; Freeman et al., 1979; Kearns, 2006). Contrary to this notion, other experiments have shown that clustered-lattice structures that reinforce a choice from various direct links but do not have a hierarchy are particularly effective in these cases (Centola, 2010). These two views may be reconciled by expecting that clustered networks are particularly successful for difficult tasks, whereas hierarchical structures are particularly successful in simple group coordination (Burt et al., 2021; Shaw, 1954). To corroborate this assumption, systematic studies with large sample sizes of groups are necessary.

One way task difficulty may vary is the divergence of individual group members' opinions about the preferred outcome due to individual dispositions (Johnstone and Manica, 2011), private information (Couzin et al., 2005), or personal needs (Rands et al., 2003), even if all individuals still desire to coordinate for group consensus. Such *opinionated individuals* within groups can be biased towards a particular decision outcome and seek to sway the group in their preferred direction. However, whether opinionated individuals manage to exert influence depends not only on their own behaviour, but also on the group's composition: Opinionated individuals who are themselves biased towards a decision outcome can—depending on their network position (Burt et al., 2021; Freeman et al., 1979; Valente and Fujimoto, 2010)—distort information flow within communication networks and ultimately affect group outcomes by selectively deciding which information to absorb (Bakshy et al., 2015) and which to transmit (Giese et al., 2020; Moussaïd et al., 2015). Moreover, indifferent individuals without a preference for a decision outcome can dampen the influence of a strongly-opinionated minority and shift the balance of power toward the numerical majority (Couzin et al., 2011).

Understanding how these complex group behaviours arise also requires uncovering the local decision-making rules of individuals. Previous work has suggested that individuals may respond to the behavior of the perceived majority, while also showing a tendency toward stability or "inertia" in their decisions (Kearns et al., 2009; Stewart et al., 2019; Ye et al., 2021). However, further empirical investigation is required to reveal the rules individuals use under different scenarios and when they hold different preferences. Moreover, agent based simulations of individual decision making rules within the communication network (Epstein, 1999) are useful to evaluate how well these rules describe both individual (Goldstone and Janssen, 2005) and group behaviour.

**Present study**

We combine online experiments with a multi-scale analytical approach to investigate how the dynamics and outcomes of consensus decisions are influenced by both the communication network structure and the presence of opinionated individuals in groups. Similar to early network coordination experiments (Burt et al., 2021; Freeman et al., 1979), we limited the group size to seven individuals. While operating with small networks imposes some limitations including less possible variation in network structure, it also comes with the key benefit of a larger sample size enabling higher statistical power. To disentangle the effects of network features and unfolding opinion dynamics, we experimentally varied the number of network links (14 vs. 8 links) and network structure (random vs. systematic) across three scenarios differing in the degree of disagreement among participants (coordination vs. leadership vs. conflict). As systematic network structures, we chose a clustered-lattice structure (14 links; see Shaw, 1954) and a specific hierarchical dumbbell structure to emulate a small-world structure (8 links; see Valente and Fujimoto, 2010; Shaw, 1954; Burt et al., 2021; Kearns et al., 2009) for comparison with their random counterparts. We postulated that a single opinionated "leader" can to some degree centralize and thus orchestrate coordination efforts (Fitch and Leonard, 2013; Kearns, 2006; Valente and Fujimoto, 2010) so that networks in the leadership scenario would form a consensus faster. In contrast, conflicts of interest between two opinionated individuals should slow down consensus formation. Furthermore, we expected that consensus formation would be generally faster in networks with more communication links (e.g., Centola, 2010). Regarding the systematic networks, we hypothesized that clustered-lattice and dumbbell networks would outperform their random counterparts in coordination and leadership scenarios (Centola, 2010; Kearns, 2006), but the hierarchical dumbbell network would cease to do so for the conflict case (Shaw, 1954).

Concerning the outcome of the group consensus, leaders should also successfully sway groups to converge on their private target colour. Generally, we expected that opinionated individuals (leaders, opponents) on more central network positions would exert more influence over decision outcomes across scenarios (Burt et al., 2021; Freeman et al., 1979; Kearns et al., 2009; Valente and Fujimoto, 2010). Please refer to https://osf.io/ubgqe for all preregistered hypotheses.

In addition to group-level dynamics and outcomes, we also investigated the rules governing *individual-level decisions* of both indifferent and opinionated group. Based on these results, we tested via agent-based simulations which local individual-level rules are sufficient to explain *both* individual decision-making and the observed dynamics at the group level.

## Materials and methods

### Participants

Participants were recruited online on the crowd-working platform *Amazon Mechanical Turk* (MTurk) and received a fee of 1.50 USD for participation. Across conditions, participants on average earned a performance bonus of 0.43 USD ($SD$ = 0.25 USD) after completion and needed about 10 min to complete the whole study. A total number of 2,394 unique participants in 342 networks started the main task (52.80% self-reported as female, $M_{age}$ = 35.78 years, $SD_{age}$ = 11.11 years).

### Procedures

To generate behavioural data on consensus decision-making, we developed a networked colour coordination paradigm (Kearns et al., 2009; Stewart et al., 2019) on *Amazon Mechanical Turk*. Groups of 7 individuals were incentivized to unanimously coordinate on a joint outcome. At the start of the task, each player was randomly assigned to a group of 7 players (network) and a position in the group (a node in the network). All players were instructed to form a group-wide consensus by unanimously selecting the same out of two colours (blue vs. yellow button) within 50 rounds. It was explained to participants that the “colour coordination game” is guaranteed to be solvable, but that all players must work together. We incentivized all participants in a group to reach a consensus as fast as possible via a monetary bonus. If the network converged on one colour, everyone received a bonus payment which started at 0.50 USD and linearly decreased at 0.01 USD increments per round down to 0.01 USD in round 50. Fig. 1 displays a screenshot of the task. If the network members did not reach consensus within the 50 rounds, participants received the participation payment of 1.50 USD only. If a participant dropped out of the study, we stopped the whole group.

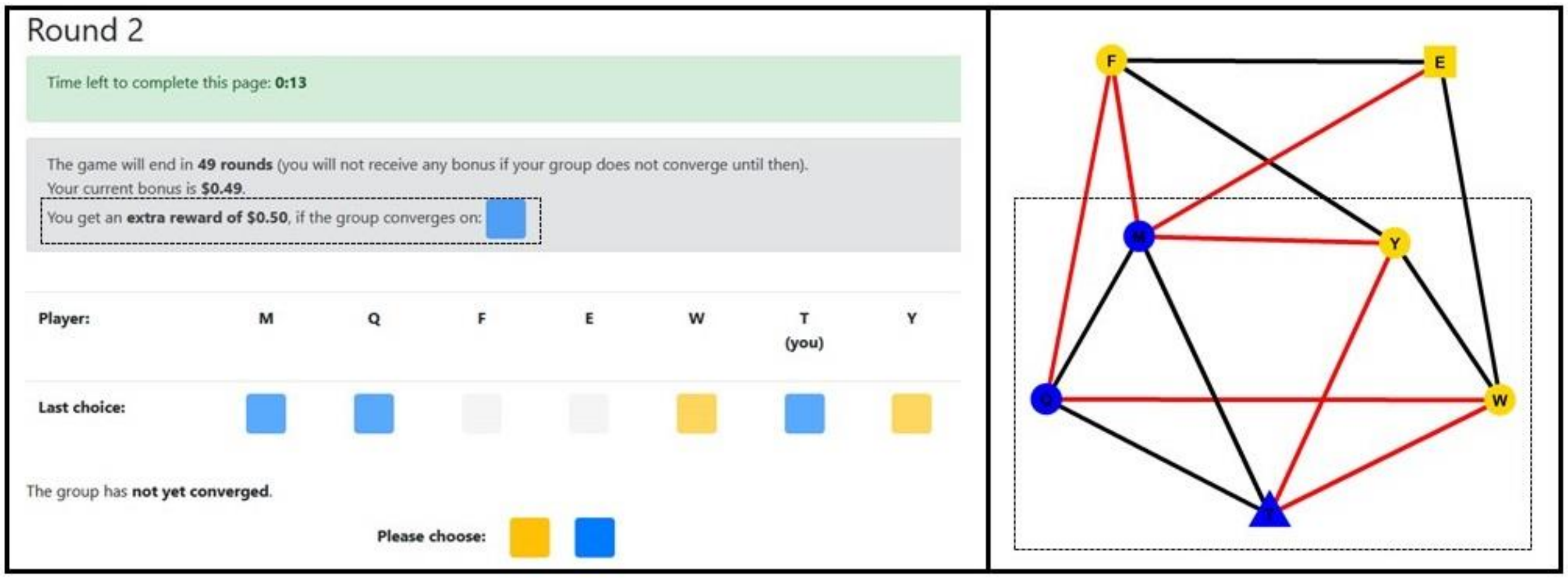

**Figure 1. Left: Screen display of the colour coordination task**. A horizontal array of coloured boxes represents the choices made in the last round. The choices of players not neighbouring in the network are greyed out. In the example, players *M, Q, W,* and *Y* are invisible to focal player *T*, who has been assigned the target to sway the group to blue. Only individuals incentivized to be opinionated about a specific target colour in the *leadership* and *conflict* scenario saw the information about the extra reward and target colour (dashed box). **Right: The underlying network of this example is an Erdős-Rényi random structure with 14 links in the conflict scenario.** Node colours represent players' choices in the last round, and node shapes represent their role, respectively. Focal player *T* (triangle) is incentivized to lead the group to "blue", whereas player *E* (square) is incentivized to lead the group to "yellow". All other players (circles) are indifferent. Red links between nodes represent local conflict, black links represent local colour convergence. The black dashed frame represents the neighbours visible to focal player *T*. Note that the network structure (right panel) was NOT visible to participants, who only saw the image depicted on the left on their screens.

## Design

We manipulated the *scenario* for consensus formation by varying the distribution of indifferent and opinionated individuals through incentivizing up to two individuals within the networks to prefer a specific colour (i.e., to be *opinionated*). In the *coordination scenario*, all seven network members were incentivized to coordinate as fast as possible while being indifferent about the outcome colour. In the *leadership scenario*, one randomly selected individual (the "leader") was additionally informed that they would receive a bonus of 0.50 USD if the network converged on a specific colour. Similarly, we randomly selected two individuals (the "opponents") with conflicting target colours for the *conflict scenario*.

Independently of the *scenario* for consensus formation (coordination vs. leadership vs. conflict), we varied the *number of links* in the network (8 links vs. 14 links), and the *structure* of the network (systematic vs. random) between networks (Supplementary Fig. 1). *Random* networks allow for general comparisons because network idiosyncrasies level out on expectation. We created random networks by using the Erdős-Rényi $G(n,m)$ model (Erdős and Rényi, n.d.), for which a network $G$ is sampled uniformly from the set of all possible networks with $n$ nodes and $m$ edges. We ensured that the network was fully connected and matching the respective fixed number of links (8 links vs. 14 links). In addition, we instantiated two theoretically and practically interesting *systematic* networks with different propensities for polarization. First, we connected individuals in a hierarchical structure with 8 links resembling a dumbbell. In this network, individuals are connected in two sub-groups that are linked through a single node to each another. Second, we connected individuals in a lattice-like structure with 14 links. This network is characterized by many short-range connections and few long-range connections between individuals. This structure is "egalitarian" in the sense that none of the individuals has a positional advantage over communication channels.

The study followed all relevant ethical regulations of [name removed for double blind review] regarding psychological research with human subjects. The study is non-invasive, and any other obvious issues concerning a threat to human health, well-being and dignity (including, e.g., deception protocols) do not apply. The study complied with ethical practices including full information, informed consent, and the option to cease participation at any point in the study, even though it was exempt from review at the Institutional Review Board of [name removed for double blind review].

## Data analyses

### *Speed of consensus formation (Survival analyses)*

To analyse the time dynamics of consensus formation on the network-level, we employ Cox Proportional Hazard survival regressions for a total of $N$ = 342 networks. The dependent variable was the network's *time-until-consensus.* We included our experimental factors as predictors: *scenario* (coordination vs. leadership vs. conflict; dummy-coded with *conflict* as reference), *number of links* (8 links vs. 14 links; effect-coded with *8 links* as reference), and network *structure* (systematic vs. random; effect-coded *random* as reference), as well as all interactions between them. We included networks that dropped out of the study as right-censored data.

Please note that *systematic networks* were either a "dumbbell" structure (for the factor level *8 links*) or a lattice-like structure (*14 links*). Therefore, our design does not fully cross all treatments (e.g., it is impossible to instantiate a 14-link dumbbell structure while keeping the number of nodes constant). However, different dynamics caused by different factor level combinations are explicitly contained in the models and picked up by interaction effects. For presenting the results, we stick to the conventional "bottom-up" approach to first report main effects and then move on to interactions and use *estimated marginal means* for unpacking comparisons.

We use *estimated marginal means,* sometimes also referred to as *predicted* or *model-based means*, for testing and reporting model effects. Estimated marginal means allow to express the mean response (as well as standard error and confidence interval) for each level of a categorical variable while statistically adjusting for any other variables in the model. Estimated marginal means can therefore support interpreting model effects, especially in the presence of interaction terms in the models. Please note that effects reported as estimated marginal means in the results section may differ from the regression coefficients reported in tables – even if they are based on the same models. For general effects of *number of links*, we focus on the two random networks. When investigating the effects of *systematic* structures, we compare the dumbbell and lattice-like network with their respective random counterparts (and not with each other).

To test and control for selective attrition, we conducted survival analyses predicting the *time-until-dropout* by our experimental conditions and treated converged networks as right-censored data. Attrition trends were statistically indistinguishable between conditions, *logrank*-$\chi^2 = 16.7$, $df = 11$, $p = 0.10$. It is therefore highly unlikely that selective attrition compromises our conclusions (see Supplemental Material for further details on recruitment).

**Influence of opinionated individuals on group outcomes**

We examined the influence of opinionated individuals on group outcomes by comparing the probability of different group outcomes (network converged vs. deadlock vs. dropout) across scenarios via $\chi^2$-tests of association or binomial tests, respectively. In addition, we examined whether the network converged on the colour option preferred by the opinionated individuals in the leadership scenario.

***Individual decision-making rules (Regression analyses)***

To statistically explore how indifferent individuals decide to pick a side and how opinionated individuals decide to give up their private target for the sake of a group consensus, we analysed a total of $N = 13{,}550$ decisions of indifferent individuals without an incentivized target colour and a total of $N = 2{,}599$ decisions of opinionated individuals with an incentivized target colour (both leaders and opponents). Data were pooled across scenarios for this analysis because individuals did not know which scenario they were in and should therefore not have systematically altered their behaviour depending on the scenario. We excluded the very first choice of participants. Furthermore, we only included choices in which all group members made a decision in the preceding round.

We modelled the probability that an indifferent individual in a network selected one colour over the other by fitting binomial generalized linear mixed models. The dependent variable was the participant’s choice of the blue option over the yellow option (reference category). We included as fixed-effects (i) the round number (*time*), (ii) the participant’s own choice in the last round (*inertia*), (iii) the proportion of blue choices in the participant’s visible network neighbourhood in the last round (*local majority*), (iv) the proportion of switches between colours by the visible network neighbours from the next-to-last to the last round (Suzuki et al., 2015, local stability), (v) the proportion of blue choices in the whole network in the last round (*global majority*), and (vi) all interactions between the participant’s local information and their actions in the last round (ii-iv) up to a four-way interaction. We modelled as random effects (a) intercepts for participants (to allow for individual initial colour preferences), (b) participant slopes for round number (to allow for different individual time trends), and (c) intercepts for networks (to allow for different starting conditions on the network-level).

We modelled individual behaviour for opinionated individuals by using a similar approach as for indifferent individuals with the following exceptions: The dependent variable was the participant’s choice of the unincentivized colour (compromising choice) over the incentivized colour option (reference category). We included the same fixed-effects but defined both local and global support as the proportion of choices to the participant’s incentivized target colour in the last round and controlled for the participant’s incentivized target colour. Random network intercepts cannot capture variation beyond individual intercepts and are obsolete because networks contain only one or two opinionated individuals. As for the survival models, we again used *estimated marginal means* to test simple slopes at most extreme values of the moderator variable (e.g., local stability set to 0 for “volatile neighbourhoods” vs. local stability set to 1 for “stable neighbourhoods”) or as indicated.

***From local behavioural rules to global dynamics (Simulations)***

While we consider the regression insightful to understand individual-level behaviour statistically, they are solely “as-if” models of the actual decision-making process of participants. To gain a deeper understanding of consensus formation our experiment, we use the regression results to instantiate a plausible rule-based account of the actual decision-making process. Thereby, we do not only move beyond statistical fitting to rule-based prediction, but can also bridge our two levels of analysis, i.e., individual, local interactions by individuals and emergent group-level behaviour: First, we simulated the decisions of individual agents embedded in networks to determine which strategies can explain *individual behaviour*. Next, we populated agents with these individual strategies on the networks from our experiment to test whether these individual rules could account for the empirically observed *group dynamics*.

**Individual level.** To test the support for different individual level rules, we simulated the behaviour of our study participants with agents employing one of a set of possible behavioural strategies (Table 1). Each agent was assigned to the network node of a participant, and could observe the empirical behaviour of the participants on neighbouring nodes (choices for the blue or yellow option, respectively) in a given round *t*. The agent then applied its assigned strategy and made a choice (blue vs. yellow) for round $t + 1$. If required by the respective strategy, the agent was also aware of its own choice in the previous round. Therefore, predictions were possible for all participants who played at least the first round ($N = 2{,}149$; $n_{indifferent} = 1{,}844$, $n_{opinionated} = 305$). We analysed the proportion of correctly predicted choices per strategy across all participants and used bootstrapping with 1,000 resampling runs to generate 95% confidence intervals around the estimates.

**Group level.** To evaluate how well individual-level rules captured group-level consensus dynamics, we populated the network from the experiment with artificial agents and simulated the behaviour of all agents forward in time to generate network-level predictions. Each agent was initialized with the actual first choice by the participant on the respective node, the choices of the observed neighbours, and the target colour of the node (if the participant was incentivized to be opinionated). The agents then subsequently made choices (blue vs. yellow) according to their assigned strategy on a round-by-round basis until the network converged or the time limit of 50 rounds was reached without a consensus. For each network we thus arrived at a prediction for the time-until-consensus (1–50 rounds) and the outcome of the consensus formation (blue vs. yellow vs. not converged). We analysed the proportion of correct outcome predictions (again using bootstrapping to generate confidence intervals) and the absolute error for predicted minus observed time-until-consensus per simulation run across all networks as measures of interest. We included participants who dropped out of the study prematurely and predicted individual-level decisions until the round before drop-out, resulting in a total of $N$ = 342 networks. For the networks affected by drop-out, we predicted the majority colour in the round before drop-out and treated the time-until-consensus as a censored observation, respectively.

To check that our simple deterministic strategies are robust to some degree of randomness in individual choices, we ran all simulations with a noise parameter $\eta$ that governed the probability that an agent would not strictly adhere to its assigned strategy but would choose a colour randomly instead. We ran simulations for the parameter values $\eta$ = 0, 0.01, 0.1, and 0.3, where $\eta$ = 0 means that an agent strictly obeys the assigned strategy, whereas $\eta$ = 0.3 means that the agent will, with a 30% probability, revert to selecting a colour uniformly at random. To stabilize results, we re-ran all simulations 500 times per participant and network and averaged the predictions across these simulation runs per agent and network, respectively.

To test if the simulated data could account for the patterns of variation across conditions and network types we observed in our behavioural data, we also conducted Cox Proportional Hazard Regressions for each of the 500 simulation runs and used the same model specifications as for the behavioural data. Examining the correlation between the Schoenfeld residuals and survival time revealed that the proportional hazards assumption does not hold for the dumbbell-shaped network in most simulation runs. This indicates that the dumbbell network structure does not only influence the chance of convergence, but also the timing of periods of high hazards for convergence. We therefore compared the distribution of hazard estimates across all simulations at the 5%, 50% (median) and 95% quantiles, respectively, with the hazards from the empirical data and their confidence bounds.

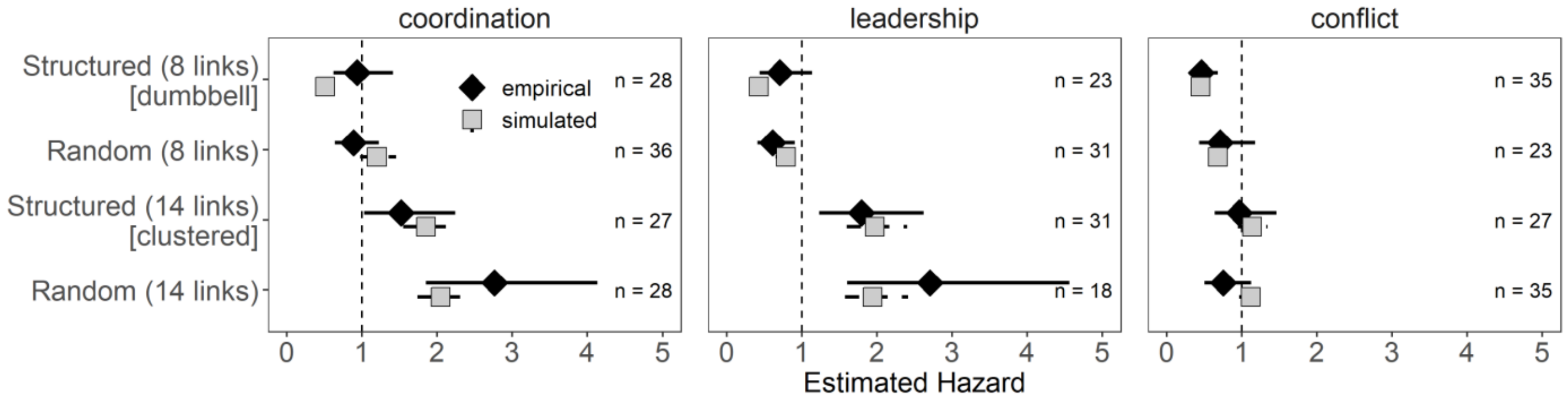

**Figure 2. Convergence speed of groups on consensus (Cox Proportional Hazard estimates), by *scenario* (coordination vs. leadership vs. conflict), *number of links* (14 links vs. 8 links), and *structure* (systematic vs. random).** Higher estimated hazard coefficients indicate faster convergence. Model fits were estimated using *N* = 342 networks (*n* denotes the number of networks per condition); symbols represent the hazard coefficient for empirical results (black diamonds) and the median hazard across all simulations (grey squares), respectively; error bars represent 95%-confidence intervals on coefficients for empirical results and the 5% and 95% quantiles of coefficients across all simulations for simulated results. The overall speed of consensus formation was lower in the conflict scenario (right panel) where two individuals in the network had contradicting preferences, as compared to the other two scenarios (left and centre panels). Networks with more links were overall faster to converge (compare coefficients for 8 links vs 14 links across all panels). However, the speed advantage of more links did not occur in the conflict scenario. By and large, we recovered the empirical trends in the data (black diamonds) with our agent-based simulations (grey squares).

## Results

### Experimental effects on the group level

#### *Speed of consensus formation (Survival analysis)*

The speed with which groups came to consensus was affected by the *number of links* in the underlying communication network, LR-$\chi^2 = 34.16$, $df = 1$, $p < 0.001$. In accordance with our predictions, networks consisting of 14 links converged faster than those consisting of 8 links ($b_{14\ vs.\ 8\ links} = .808$, $p < 0.001$, $HR = 2.260$ [95% CI: 1.72, 2.96]). Because network idiosyncrasies level out on expectation, comparisons between random networks provide the appropriate test for the general influence of different numbers of links. As such, this facilitating effect of more links can be specifically found also when only comparing the number of links for random networks ($b_{14\ vs.\ 8\ links\ |\ random} = .896$, $p < .001$, $HR = 2.45$ [95% CI: 1.96, 3.56]) and therefore seems to be rather general across different network structures.

In addition, the given *scenario* also affected consensus speed, LR-$\chi^2 = 22.30$, $df = 2$, $p < 0.001$. As expected, conflicting preferences resulted in slower consensus formation compared to both pure coordination ($b_{coor\ vs.\ confl} = 0.670$, $p_{adj} < 0.001$, $HR = 1.95$ [1.35, 2.83]), and leadership, ($b_{lead\ vs.\ confl} = 0.543$, $p_{adj} = 0.003$, $HR = 1.72$ [1.16, 2.56]). Contrary to our expectations, having a single "leader" did not speed up consensus compared to situations in which all group members were indifferent ($b_{lead\ vs.\ coord} = -.128$, $p_{adj.} = 1.000$, $HR = 0.88$ [0.61, 1.27]). In a follow-up analysis to explore the role of the initial majorities, we compared the coordination and leadership scenario by adding a variable with 3 levels capturing whether (1) the group's choices in the first round supported the opinionated individual, (2) opposed it, or (3) if there was no opinionated individual (pure coordination, as reference) (*LR*-$\chi^2 = 10.32$, $df = 2$, $p = 0.006$). We found that, if opinionated leaders had to override an initial, spontaneous majority against them, convergence came with a speed penalty compared to both initial support ($b_{con\ vs.\ pro} = -0.763$, $p_{adj} = 0.045$, $HR = 0.47$ [0.26, 0.09]), and the pure coordination scenario ($b_{con\ vs.\ coor} = -0.535$, $p_{adj} = 0.003$, $HR = 0.59$ [0.35, 0.99]).

Regarding our predictions for the speed of convergence of systematic networks (clustered-lattice and dumbbell), we did not find that network *structure* on average influenced convergence speed (LR-$\chi^2$ = 0.40, $df$ = 1, $p$ = 0.52), indicating that overall the two *systematic* networks do not come to consensus significantly faster than *random* ones based on the Erdős-Rényi $G(n,m)$ model (Erdős and Rényi, n.d.): Relative to their random counterparts with the same number of links in each of the scenarios, the systematic structures *did not* perform differently in the coordination ($b_{clustered\ vs.\ random\ 14\ |\ coordination}$ = -0.599, $p_{adj}$ = 0.243, $HR$ = 0.55 [0.25, 1.19]; $b_{dumbbell\ vs.\ random\ 8\ |\ coordination}$ = 0.056, $p_{adj}$ = 1, $HR$ = 1.06 [0.51, 2.20]), leadership ($b_{clustered\ vs.\ random\ 14\ |\ leadership}$ = -0.411, $p_{adj}$ = 1, $HR$ = 0.66 [0.27, 1.61]; $b_{dumbbell\ vs.\ random\ 8\ |\ leadership}$ = 0.149, $p_{adj}$ = 1, $HR$ = 1.16 [0.49, 2.74]) or conflict scenario ($b_{clustered\ vs.\ random\ 14\ |\ conflict}$ = 0.248, $p_{adj}$ = 1, $HR$ = 1.28 [0.57, 2.90]; $b_{dumbbell\ vs.\ random\ 8\ |\ conflict}$ = -0.440, $p_{adj}$ = 1, $HR$ = 0.64 [0.26, 1.60]).

Interestingly, the speed advantage of more links seemed to be marginally dampened under conflict of interest (*scenario* × *number of links* × *structure*, LR-$\chi^2$ = 5.37, $df$ = 2, $p$ = 0.069). Again, comparisons only considering random networks provide the appropriate test of how the consensus formation speed is generally affected by numbers of links in the different scenarios. These tests showed that the effect of more links for faster convergence decreased for random networks with increasing levels of conflict (*scenario* × *number of links* | *random*, LR-$\chi^2$ = 8.92, $df$ = 2, $p$ = .012; Fig. 2): While random networks with 14 links outperformed their counterparts with 8 links both in the leadership scenario ($b$ = 1.493, $p_{adj}$ < 0.001, $HR$ = 4.45 [2.25, 8.81]), and in the coordination scenario ($b$ = 1.140, $p_{adj}$ < 0.001, $HR$ = 3.13 [1.82 5.39]), the speed advantage of more communication links on consensus formation vanished under conflict ($b$ = 0.056, $p_{adj}$ = 0.87, $HR$ = 1.06 [0.54, 2.08]).

Summing up, we found that groups under local communication came to consensus faster if there were more communication channels (network links) and if opinionated individuals were absent. A single opinionated individual slowed down reaching a consensus if opposed by an initial (spontaneous) majority. Coordinating a consensus was slowest under conflict of interest—in this case, even higher network connectivity did not speed up consensus formation.

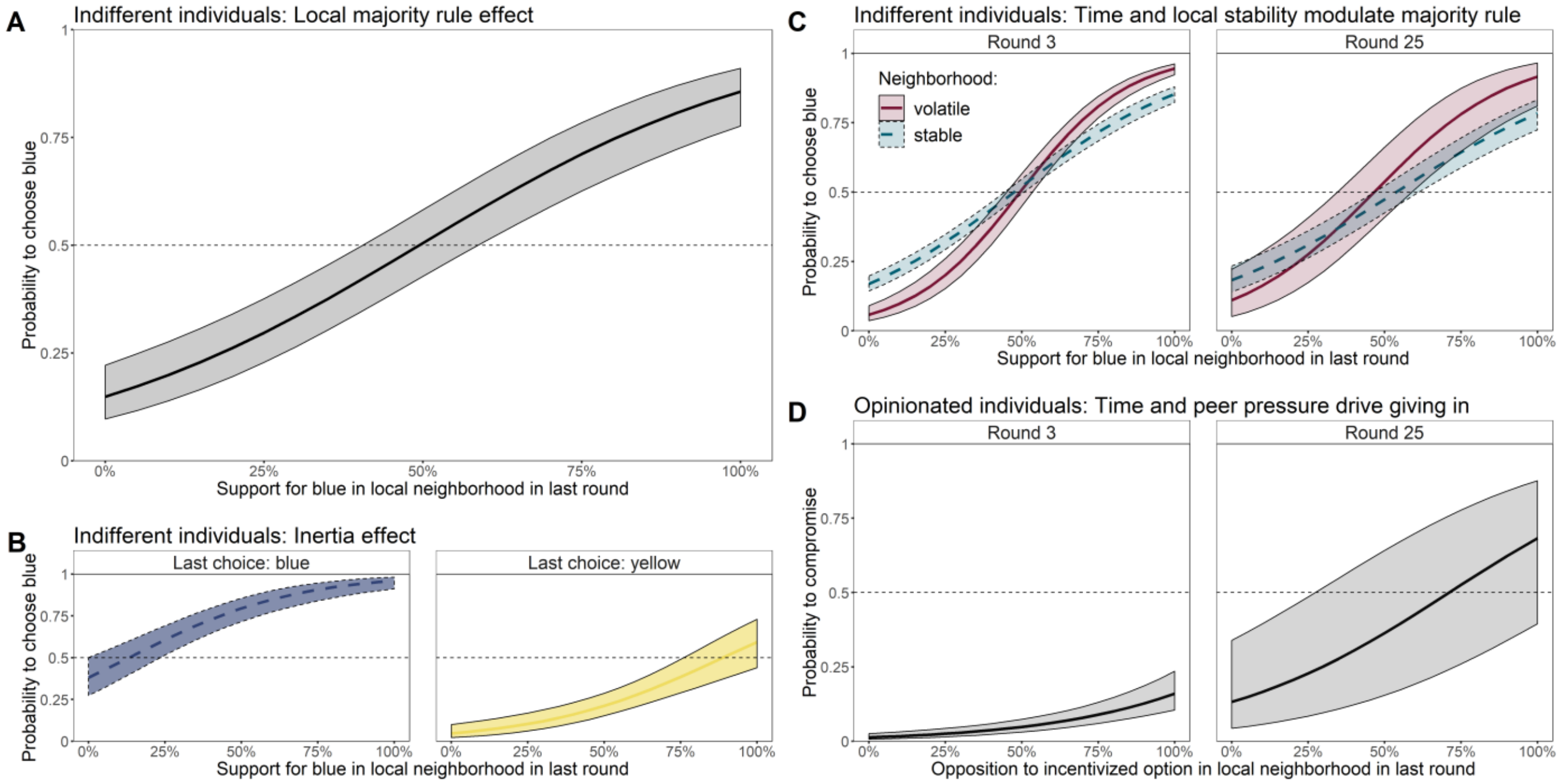

**Figure 3. Probabilities for individual decisions as predicted by regression models.** Results rely on a total of 13,550 individual-level decisions of indifferent individuals and 2,599 of opinionated individuals, respectively. Lines represent model predictions and ribbons represent 95%-confidence intervals. (**A**) Indifferent individuals followed the majority colour in their local neighbourhood (local majority rule effect). Plot shows probability that an individual chose blue as a function of the % of its neighbours choosing blue in the last round. (**B**) Despite the local majority rule effect, indifferent individuals were hesitant to switch away from their last colour choice (inertia effect), with the probability to choose blue overall higher when an individual's previous choice was blue (left panel) than when it was yellow (right panel). (**C**) The influence of the local majority rule was stronger (i.e., steeper curve) in volatile (red ribbon; local stability set to 0) compared to stable (teal ribbon; local stability set to 1) neighbourhoods at the onset of the task (left panel), but not later on (right panel). (**D**) Opinionated individuals were more likely to compromise under local opposition (peer pressure) and after some time had passed (time pressure). Plots show model-predicted probabilities that an individual deviated from its incentivized colour as a function of the % of its neighbours choosing the opposite colour in the previous round, both early in the game (left panel) and late in the game (right panel).

***Influence of opinionated individuals on group outcomes***

Regarding the outcome of the consensus dynamics, we found that a single, opinionated individual tended to dictate which colour the group chose, thus effectively exerting leadership. The speed and success rate of leadership, however, were both independent of network structure. Out of the 103 leadership networks, 75.7% converged (95% Clopper-Pearson-CI [66.3%, 83.6%]). Of the converged networks, most converged on the leader's target colour ($prob_{converged\ on\ leader\ color\ |\ converged}$ = 0.73 [0.62, 0.82]). Instances of deadlock (in which the group did not achieve a consensus within 50 rounds) were rare ($prob_{deadlock}$ = 0.05 [0.02, 0.11]), and drop-out (in which games were terminated due to individuals abandoning the online study) was comparatively low (Arechar et al., 2018) ($prob_{drop-out}$ = 0.19 [0.12, 0.28]). There was no association between network features (number of links and structure) and the outcome of consensus formation under leadership ($\chi^2$ = 5.51, $p_{simulated}$ = 0.81), and the leader's relative number of direct neighbours (degree centrality) did not predict consensus speed ($LR\text{-}\chi^2$ = 0.01, $df$ = 1, $p$ = 0.91; see also Supplementary Table 2).

Even under conflicts of interest, most of the 120 networks in this scenario converged and hence successfully averted deadlocks ($prob_{converged}$ = 0.65 [0.56,0.73]; $prob_{deadlock}$ = 0.03 [0.005, 0.071], $prob_{dropout}$ = 0.33 [0.24, 0.42]). Again, the outcome of consensus formation (convergence vs. deadlock vs. drop-out) was independent of the network's features (number of links and structure; $\chi^2$ = 5.04, $p_{simulated}$ = 0.55). Contrary to our predictions, we did not find a direct effect of difference in degree centrality between opponents on convergence speed (Supplementary Table 3), nor was there an increased chance that an opponent with higher (relative) degree centrality prevailed over their rival in the network (42.42% [25.48%, 60.78%], $p_{binomial\ test}$ = 0.49).

Summing up, we found that single, opinionated individuals exerted leadership by swaying groups to their preferred outcome. The speed and success rate of leadership, however, were independent of network structure and the individual's position in the network, across the range of network structures we tested. The presence of two opposing, opinionated individuals slowed down consensus formation significantly, but in most cases, groups were still able to avoid deadlock.

**Individual decision-making rules (Regression analysis)**

We found that the decisions of indifferent individuals were driven by an “inertia effect”, such that participants tended to stay with their previous choice rather than switching colours, as well as a tendency to synchronize to the local majority colour. This “follow-the-majority rule” was stronger in volatile compared to stable neighbourhoods (with stability defined as the local switching rate of observed neighbours between rounds) at the onset of the task, but not later on.

More precisely, the greater the fraction of an individual’s local neighbours that chose a given colour, the higher the probability that the individual would also choose that colour in the subsequent round (Fig. 3A). This *local* majority rule explains choice behaviour above and beyond the *global* proportion of choices for a given colour in the network (LR-$\chi^2$ = 1163.80, $df = 1$, $p < 0.001$, $b_{overall\ majority\ slope} = 3.53$, 95%-CI [2.71, 4.35]). Despite the normative influence of the local majority, participants were generally hesitant to move away from their previous colour choice, indicated by an auto-regressive or “inertia” effect (Fig. 3B; LR-$\chi^2$ = 2150, $df = 1$, $p < 0.001$, $b_{last\ choice} = 2.67$ [2.11, 3.22]). The strength of the local follow-the-majority rule was furthermore modulated by the majority’s local stability and the number of rounds passed (*local majority* × *local stability* × *time*, LR-$\chi^2$ = 9.78, $df = 1$, $p = 0.002$; predictions of the best-fitting model are presented in Fig. 3C, the models are disclosed in Supplementary Table 4). At onset, the local majority rule was stronger in volatile neighbourhoods than in stable neighbourhoods (volatile: $b_{majority\ slope} = 5.64$, [4.91, 6.38]), stable: $b_{majority\ slope} = 3.35$, [2.99 ,3.71], $p < 0.001$). This pattern of behaviour can be interpreted as exploring choice options in stable situations and displaying consistency in locally volatile situations to facilitate the emergence of consensual options. In the middle of the task (round 25), the local majority rule was still influential, but not significantly different between stable and volatile environments (volatile: $b_{majority\ slope} = 4.47$, [3.05, 5.90], stable: $b_{majority\ slope} = 2.78$, [2.30, 3.27], $p = 0.196$; the full interaction also including a colour bias is displayed in Supplementary Fig. 2).

For opinionated individuals, we found that both local opposition ($LR\text{-}\chi^2 = 55$, $df = 1$, $p < 0.001$) and time pressure ($LR\text{-}\chi^2 = 17$, $df = 1$, $p < 0.001$) impact their propensity to give in (Fig. 3D): Opinionated individuals were less likely to switch to their non-preferred option the more a local majority supported their desired outcome ($b_{local\ support} = -2.67$, 95%-CI [−3.30, −2.04], $OR = 0.069$). In addition, they were more likely to switch the more time had already passed ($b_{round} = 0.112$, [0.05, 0.17], $OR = 1.12$). Seemingly, forgoing some payoffs due to a bad deal was preferred to foregoing all payoffs because of deadlock. In contrast to our findings for indifferent individuals, for opinionated individuals the stability of local majorities did not moderate the influence of the local majority ($LR\text{-}\chi^2 = 0.23$, $df = 1$, $p = 0.63$). The models are fully disclosed in Supplementary Table 5.

In summary, by probing which local information predicts individual decisions, we found that indifferent individuals were hesitant to switch colours ("inertia" effect), but overall displayed behaviour consistent with a follow-the-majority heuristic in the local neighbourhood that is sensitive to the local stability of group preferences. If local majorities were stable but did not result in a global consensus, individuals were more likely to deviate from that local norm. While opinionated individuals were sensitive to time and local norms, they did not change their behaviour depending on the temporal stability of such norms—in other words, they tried to get their will by insisting on their preferred option.

**From local behavioural rules to global dynamics (Simulations)**

***Individual level***

In our individual agent-based simulations, a local follow-the-majority rule that reverts to inertia in case of ties best describes the behaviour of *indifferent individuals* across the set of rules we tested (Table 1), correctly predicting approximately 3 out of 4 individual decisions (Fig. 4A). Pure inertia and guessing (choosing an option at random) rules perform poorly at predicting the decisions of indifferent individuals, indicating that completely ignoring social information does not account for actual behaviour. While copying a randomly selected neighbour performs above chance-level, it is outperformed by simple majority-based rules. This indicates that a tendency for social conformity explains the data better than simple copying. Likewise, strategies relying on unanimous majorities perform slightly worse than strategies relying on simple majorities, indicating that unanimity in the neighbourhood is too strict a requirement for adapting one's behaviour. The rank order between models is robust even when incorporating up to 30% of noise in agent decisions.

Among the simulated strategies for *opinionated individuals* (Supplementary Fig. 3), pure inertia performs very well (as was to be expected from the regression results) and thus serves as a powerful upper benchmark for the other strategies. The second-best strategy describing opinionated individuals is a follow-a-unanimous-majority-rule that relies on the last choice (inertia) in case there is no unanimous majority. This finding reflects the importance of peer pressure found in the regression analyses for triggering compromising choices if individuals hold strong private preferences about the group outcome. While unanimity proved to be too strict a criterion for indifferent individuals, a simple majority apparently is too weak as peer pressure for opinionated individuals to give up private outcomes. Again, the rank order between models is robust to incorporating noise levels of up to 30%.

**Table 1.** Strategies used in agent-based simulations

| No. | Strategy | Description |
|---|---|---|
| 1 | Guessing | Choosing a colour option uniformly at random. |
| 2 | Copy random neighbour | Randomly selecting one neighbour and copying its choice from the last round. |
| 3 | Simple majority rule with guessing | Follow local majority from last round; if there was none, guess. |
| 4 | Simple majority rule with inertia | Follow local majority from last round; if there was none, repeat own last choice. |
| 5 | Unanimous majority with guessing | If *all* local neighbours agreed on a colour in last round, follow them; guess otherwise. |
| 6 | Unanimous majority with inertia | If *all* local neighbours agreed on a colour in last round, follow them; repeat own last choice otherwise. |
| 7 | Pure inertia | Always stick to first choice and never alter irrespective of what others are doing. |

*Notes:* This set of strategies covers a spectrum from purely random behaviour over variants of local majority-based rules (Gigerenzer and Gaissmaier, 2011; Hastie and Kameda, 2005; Hertwig et al., 2012; Suzuki et al., 2015) to the extreme of pure inertia. We varied the specifications of majority rules regarding (1) how strongly agents take the local majority into account (simple vs. unanimous majority) and (2) how the agent deals with uninformative majorities (ties in the local neighbourhood). A “pure inertia” strategy captures that forgoing a group-wide consensus is preferred over giving up one’s preference (Stewart et al., 2019).

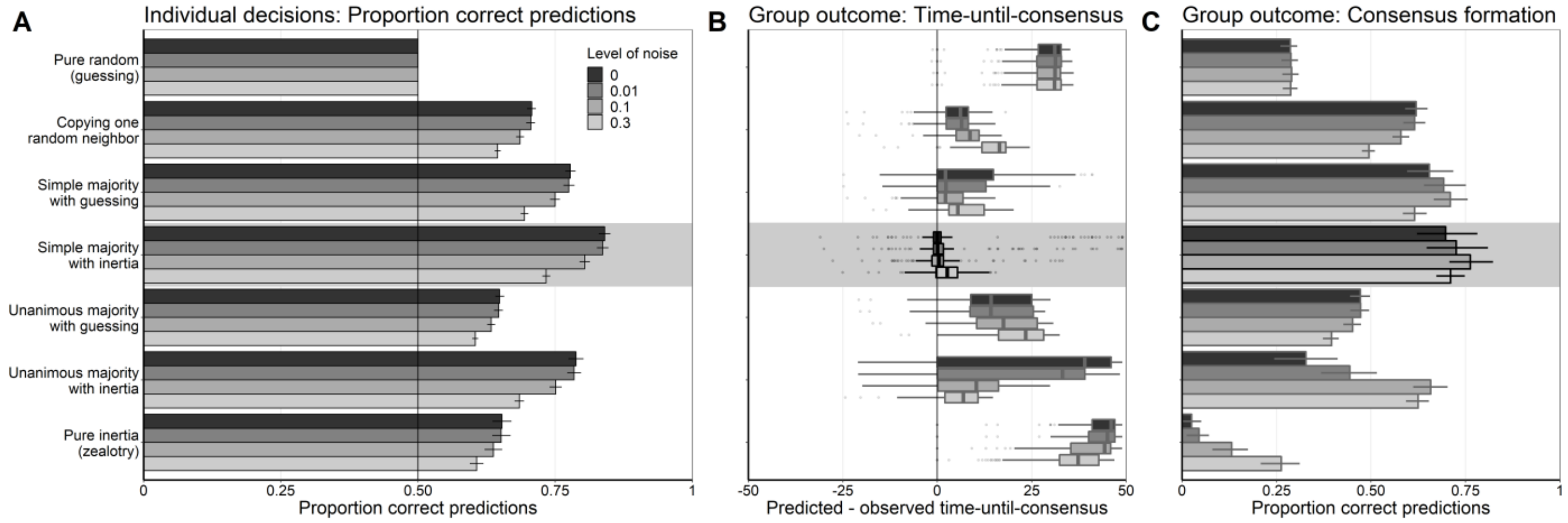

**Figure 4. Simulation results for indifferent individuals show that a majority-with-inertia rule best describes decision-making at both the individual and the group level.** (**A**) Proportion of correct predictions of individual decisions for each type of simulated agent (labels at left) under varying levels of noise (bar shading colour). (**B**) Difference between predicted time until group consensus in simulations and in empirical data. Values near zero indicate accurate predictions, whereas values above (below) zero indicate that simulations over-predicted (under-predicted) time to convergence.
(**C**) Proportion of correct predictions of group-level decision-outcome for each type of simulated agent. Bars in **A** and **C** represent the mean proportion of correct predictions across simulation runs and error lines represent 95% confidence intervals of the means based on 1000 bootstrap samples. Bars in **B** represent box plots (25th and 75th percentiles) and lines represent whiskers (extensions from 25th and 75th percentiles by 1.5 × inter-quartile range). The centre line denotes the median. Depicted results are based on the average of 500 simulation runs per participant and network, respectively. The best performing rule (majority with inertia) is highlighted by grey shading. Results for opinionated individuals are documented in Supplementary Fig. 3 and 4.

***Group level***

Group dynamics between exclusively indifferent individuals in the *pure coordination scenario* are best described by a simulated local follow-the-majority rule with inertia in case of ties and a moderate amount of noise. A local majority rule with inertia explains well both the time-until-consensus (Fig. 4B) and the group outcome (i.e., which colour was ultimately selected, Fig. 4C) in our observed networks. All other simulated strategies perform worse in accurately predicting the consensus outcome, and drastically over-predict the time until consensus. We found that incorporating a moderate amount of noise (10%) on all individual decisions leads to more accurate predictions than incorporating either higher or lower levels of noise. In sum, a majority-with-inertia rule with some noise well describes both individual-level decisions and group-level dynamics for the pure coordination scenario.

To simulate group dynamics in the *leadership* and *conflict* scenarios, we modelled indifferent agents with the best-supported strategy (majority-with-inertia rule and 10% of individual-level noise) and systematically varied strategies (Table 1) and noise-levels for opinionated agents. However, the simulations for both the *leadership* and *conflict* scenario showed that consensus dynamics are largely invariant to the exact strategy assumed for opinionated individuals (Supplementary Fig. 4). Instead, virtually all strategies for the opinionated individual provide correct predictions of the outcome of consensus formation for about 60% of all networks. In addition, the bulk of predictions for the time-until-convergence are only about 8–10 rounds off and without any clear indication of over- or under-prediction. Pure inertia, which will never compromise, is the notable exception and predicts non-convergence in all cases and thus performs poorly, unless softened by noise. Recall that we populated the agents on our empirical network and fed in the observed starting conditions before the agents "hands-off" employed their strategies. Therefore, our simulations underscore that given a network with an initial (preference-based) opinion distribution and a group of indifferent individuals applying a simple, local follow-the-majority rule with inertia, the empirically observed consensus dynamics emerge largely irrespective of the simulated individual strategy of opinionated individuals.

While the results above indicate that our simulated rules can capture both individual-level decisions and group-level dynamics, we further tested the power of these simulations to accurately predict patterns of consensus dynamics across different network structures. To do so, we used the data sets generated from the agent-based simulations and analysed the time-until-consensus on the group-level with Cox Proportional Hazard regressions (using the same model structure as was used for the empirical data). The results show that, using simulated data in the model fits, we successfully recover the main behavioural findings on consensus dynamics as observed in the empirical data (Fig. 2). In sum, our simple, deterministic rules provide reasonable predictions of *both* individual-level decision-making and the emergent group outcomes and opinion dynamics across varying network structures and preference distributions.

## Discussion

Our results demonstrate that the dynamics and outcomes of consensus decisions can be shaped both by the underlying communication network and by the preferences of individuals within groups. We replicate (Centola, 2010) that higher network connectivity improves the speed of consensus formation, but newly find that this acceleration effect does not hold under conflict of interest when two group members try to sway the group toward opposing decision outcomes. In contrast to our expectations, single opinionated individuals (“leaders”) do not speed up global consensus formation but can potentially slow down consensus-seeking groups when their preferences are initially (spontaneously) opposed by other group members. Finally, our results demonstrate that a local majority-with-inertia rule can explain both individual decisions and the global dynamics and outcomes of consensus decision-making processes.

When groups consist of only indifferent individuals, coming to consensus is fast. This speed results from amplifying random, initial leanings in the group and the absence of opinionated individuals trying to push their own agenda. Having more communication channels between individuals speeds up consensus formation. Beyond the number of links, this swiftness was robust to the exact structure of the network, at least among the limited range of network structures we investigated here. These findings epitomize how efficient decisions can emerge without global vote-taking or a central authority from solely local interactions (Conradt and Roper, 2005). Yet, the lack of findings on the role of specific structures, which can be seen in conflict with earlier findings (Centola, 2010; Kearns et al., 2009; Stewart et al., 2019), should be interpreted cautiously, as the small size of the tested networks may have rendered the compared structures functionally too similar. Another caveat on the effect of network structure results from our design in which the levels of the factor network *structure* were not fully crossed for the systematic networks (dumbbell vs. clustered-lattice). However, we *statistically* disentangled the effects of number of network links and network structure using estimated marginal means. Thereby, the comparisons between random networks remain valid even if one would want to place more caution on the findings regarding systematic networks.

In agreement with our expectations, we find that single opinionated individuals can exert effective leadership (Smith et al., 2016) and sway collective decisions to their preferred outcomes. This is well in line with theoretical models and empirical work in biology that predict that a proportion of opinionated individuals as small as 10% can strongly sway group outcomes (Couzin et al., 2011, 2005; Dyer et al., 2009). However, we also expected that the single, opinionated individual would to some degree "centralize" and thus speed up the coordination effort (Kearns, 2006). Contrary to our expectations, consensus formation under leadership was on average equally fast as under pure coordination, or could even be slower if the leader had to overrule a spontaneous, initial majority. In addition, the leader's chances to win over the group and the speed of consensus formation were independent of their position in the network and the global network structure, in contrast to some theoretical predictions (Fitch and Leonard, 2013), although it is important to note that this finding may not generalize to larger or differently structured networks and lacks the statistical power for a final verdict. Overall, these findings suggest that the speed of consensus formation is not driven by an opinionated individual that orchestrates local coordination by providing a joint goal. Rather, indifferent individuals absorb the interests of single, strongly opinionated individuals and this integration effort in the group can take time. While swiftly incorporating the preferences of opinionated individuals can minimize costs (i.e., forgone payoffs) for *both* the group and the opinionated individuals (Conradt and Roper, 2005, 2003) and thus be adaptive, strongly opinionated leaders pursuing private interests could also exploit these dynamics to the majority's detriment (Couzin et al., 2011; Stewart et al., 2019).

Consensus formation in the case of conflicting opinions takes longer than both pure coordination and leadership situations, as we predicted. Furthermore, we find that opinionated individuals are more willing to compromise and choose their non-preferred option with growing time pressure and larger peer pressure, indicating that opinionated individuals are also sensitive to local majorities. Crucially, however, the speed advantage of greater network connectivity between group members cannot be harvested in these conflict scenarios. This result suggests that attempts to accelerate the resolution of preferential conflicts of interest by increasing the quantity of communication opportunities may prove ineffective. Nevertheless, our networks typically reached consensus eventually, thus successfully avoiding deadlocks. Our study participants therefore did not behave as un-compromising zealots (Stewart et al., 2019), but seemed to rely on "tasteful stubbornness", i.e., insisting on their preferred colour although it is the minority option in the local neighbourhood (Kearns et al., 2009). While our simulations of opinionated individuals reveal that strategies incorporating more indifference or leniency in the form of noise (i.e., random choices) are better able to capture group-level dynamics than their rigid counterparts, our purely deterministic rules are undoubtedly not flexible enough to capture the nuances of real human behaviour.

Our results show that a simple "majority-with-inertia" rule well captures both the decisions of indifferent individuals and group-level outcomes in statistical and deterministic agent-based models. While the robust performance of majority-based rules (Gigerenzer and Gaissmaier, 2011; Hertwig et al., 2012) has been demonstrated previously for both human (Hastie and Kameda, 2005; Kerr and Tindale, 2004) and non-human (Conradt and Roper, 2003; Simons, 2004) animals, we show that it also prevails under settings of local communication. As crucial new additions, we demonstrate that both inertia and the stability (Suzuki et al., 2015) of local majorities (which could also be interpreted as the inertia of others) are not necessarily a nuisance but can provide valuable information to signal and gauge the consensuality of group outcomes in repeated local interactions. Generally, indifferent individuals tended to stick with their previous choices (inertia effect), thus making their actions predictable. In a given local neighbourhood this may create the stability necessary for all neighbours to mutually synchronize their choices to the prevailing majority colour. However, if majorities in the local neighbourhood are stable but do not result in global consensus, individuals are more likely to deviate from their previous choices, thus switching from inertia to exploration. Therefore, indifferent individuals seem to use the temporal stability of choice options as cues for the potential to find consensus. Locally adhering to this rule may provide an effective balance of conforming for consensus and deviating to avoid deadlock, which ultimately brings about global consensus.

We also find that both the outcome and duration of consensus formation are largely robust to the simulated behaviour of opinionated individuals if indifferent individuals follow a majority-with-inertia rule. That consensus dynamics unfolded nevertheless in a predictable way suggests that indifferent individuals in the group were largely responsible for driving the dynamics of the search for consensus (Couzin et al., 2011). While this finding could also be due to the small number of opinionated individuals, it nevertheless highlights the need for future research on group decision making to broaden its focus from the influence of opinionated individuals to the potential importance of indifferent individuals in consensus dynamics (Couzin et al., 2011), and how they consider local information and its temporal stability.

It has been repeatedly argued that submission to peers ("copying") can lead to cascades of incorrect decisions (Rahwan et al., 2014; Rendell et al., 2010), thus epitomizing "irrational" or "maladaptive" behaviour. For example, a recent study demonstrated that *zealots* (scripted agents preferring deadlock to losing) cause strongly biased information cascades that sway outcomes of group votes in networks, even if the groups are numerically at par (Stewart et al., 2019). The authors conclude that their study "provides an account of the vulnerabilities of collective decision-making" (Stewart et al., 2019). While non-compromising decision makers exist, we stress that instances of zealotry are *boundary* *conditions*. In our decision paradigm with a similar MTurk sample and incentivization scheme, we find that our participants rely on a mechanism that can harvest the "robust beauty of majority rules in group decisions" (Hastie and Kameda, 2005) and effectively and efficiently avoid deadlocks through compromising and breaking up frozen majorities. The advantage of such exploratory actions has recently been demonstrated by using scripted agents as well (Shirado and Christakis, 2017), which further highlights that zealotry is but one specific configuration of the real social world. We therefore side with previous conclusions (Conradt and Roper, 2005; Rendell et al., 2010; Sumpter et al., 2008) that negative cascades in groups can be explained as by-product of usually accurate and adaptive mechanisms for consensus decision making.

## Authors' contributions statement

All authors contributed to the conceptual development of the work. H.G. and A.S.P. designed the experiments; F.G. programmed the experiments and collected the data; F.G. analyzed the data and carried out the simulations with inputs by H.G. and A.S.P.; F.G. produced the first draft, all authors provided critical revisions and approved of the final version.

## Acknowledgements

We thank Nico Gradwohl for his help with programming the grouping mechanism for the online experiments.

## Funding

This work was supported by the *Zukunftskolleg* of the University of Konstanz (H.G. and A.S.P.); the *Graduate School of Decision Sciences* at the University of Konstanz (F.G.); the *Gips-Schüle Stiftung* (A.S.P.); and the Deutsche Forschungsgemeinschaft (DFG, German Research Foundation) [grant numbers 441541975 (H.G.) and under Germany's Excellence Strategy – EXC 2117 – 422037984].

## Contact details

Felix Gaisbauer, Social Psychology and Decision Sciences, Department of Psychology, University of Konstanz, P.O. Box 43, 78457 Konstanz, Germany

felix.gaisbauer@uni-konstanz.de

## Additional information

**Supplementary Information** accompanies this paper.

**Data Availability:** Data files as well as code for producing results, graphs, and simulations are available at https://osf.io/k9zsf.

**Declaration of interests:** None.